\documentclass[runningheads]{llncs}
\usepackage[T1]{fontenc}
\usepackage{graphicx}
\usepackage{amsmath}
\usepackage{amssymb}
\usepackage{booktabs}
\usepackage{algorithm}
\usepackage{algorithmic}
\usepackage{multirow}
\usepackage{comment}
\usepackage{subfigure}
\usepackage{color}
\usepackage{bm}

\begin{document}

\title{Pose-GuideNet: Automatic Scanning Guidance for Fetal Head Ultrasound from Pose Estimation}
\titlerunning{Guidance for Fetal Head US from Pose Estimation}
%
\author{
Qianhui Men\inst{1} \and
Xiaoqing Guo\inst{1} \and
Aris T.\ Papageorghiou\inst{2} \and
J.\ Alison Noble\inst{1}
}

\authorrunning{
Q.\ Men et al.
}
%
\institute{
Institute of Biomedical Engineering,
\mbox{University of Oxford}, Oxford, UK\\
\email{qianhui.men@eng.ox.ac.uk}
\and
Nuffield Department of Women's \& Reproductive Health, \mbox{University of Oxford, Oxford, UK}
}
\maketitle              

\begin{abstract}
3D pose estimation from a 2D cross-sectional view enables healthcare professionals to navigate through the 3D space, and such techniques initiate automatic guidance in many image-guided radiology applications. In this work, we investigate how estimating 3D fetal pose from freehand 2D ultrasound scanning can guide a sonographer to locate a head standard plane. Fetal head pose is estimated by the proposed \textit{Pose-GuideNet}, a novel 2D/3D registration approach to align freehand 2D ultrasound to a 3D anatomical atlas without the acquisition of 3D ultrasound. To facilitate the 2D to 3D cross-dimensional projection,
we exploit the prior knowledge in the atlas to align the standard plane frame in a freehand scan. A semantic-aware contrastive-based approach is further proposed to align the frames that are off standard planes based on their anatomical similarity. In the experiment, we enhance the existing assessment of freehand image localization by comparing the transformation of its estimated pose towards standard plane with the corresponding probe motion, which reflects the actual view change in 3D anatomy. Extensive results on two clinical head biometry tasks show that Pose-GuideNet not only accurately predicts pose but also successfully predicts the direction of the fetal head. Evaluations with probe motions further demonstrate the feasibility of adopting Pose-GuideNet for freehand ultrasound-assisted navigation in a sensor-free environment.

\end{abstract}

\section{Introduction}

\begin{figure*}[]
    \centering
    \includegraphics[width=1.0\textwidth]{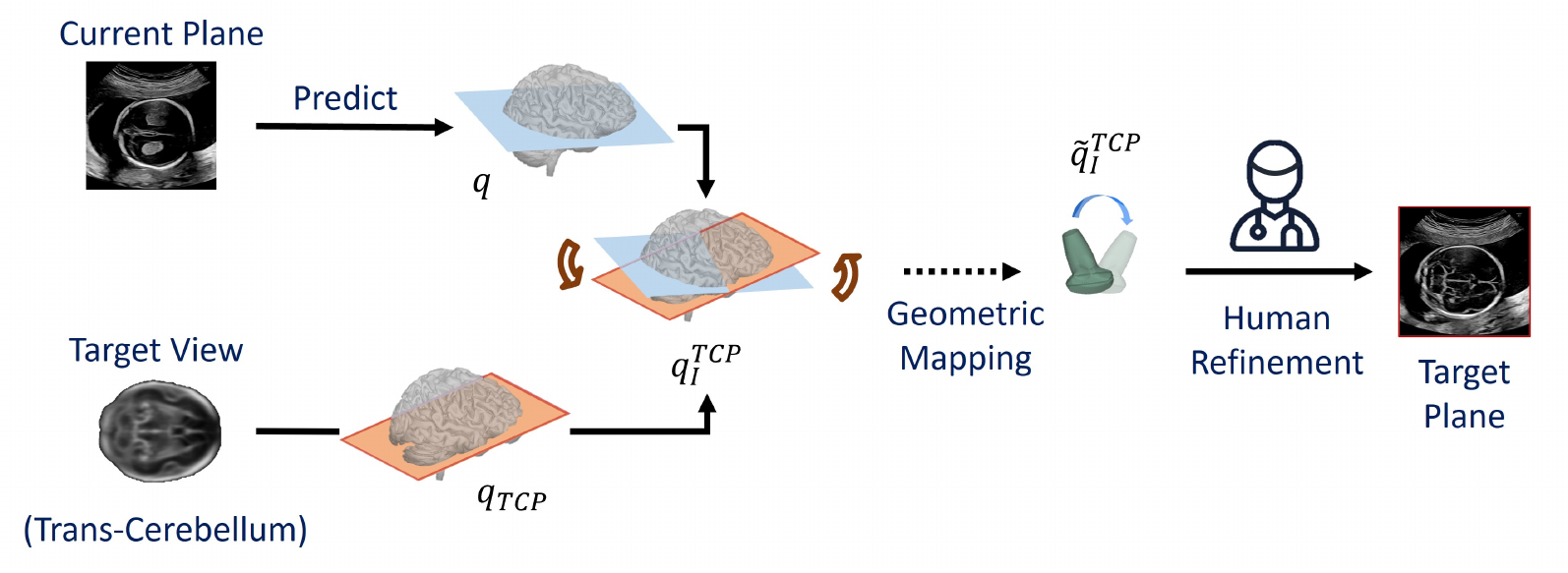}
    \caption{Principle of \textit{Pose-GuideNet}. The system provides guidance on how to move to a SP of 3D anatomy, and the corresponding probe movement would be inferred through geometric mapping from 3D anatomy to the probe coordinate.}
    \label{fig:principle}
\end{figure*}

Ultrasonography requires an operator to be proficient in 3D spatial perception to interpret the entire region of interest from the perspective of a single acoustic window. Clinical sonography workflow usually involves coarsely localizing anatomy of interest, and then ultrasound probe refinement to reach a high-quality anatomical standard plane (SP) for accurate examination. The task is extremely challenging for obstetric clinicians when examining multiple anatomies of a moving fetus~\cite{salomon2011practice}. An automatic system combining the two processes - fetal pose localization and navigation to SP (as described in Fig.~\ref{fig:principle}), would greatly reduce scanning workload as well as provide trainee sonographers and occasional users with heuristic guidance of probe manipulation to support SP acquisition.

Existing algorithms in the literature typically regard fetal SP detection as a classification problem of still ultrasound (US) images~\cite{baumgartner2017sononet,lee2021principled,sundaresan2017automated} or video clips~\cite{chen2017ultrasound,pu2021automatic}. Recently, guidance-based frameworks have been developed~\cite{droste2020automatic,men2022multimodal,men2023gaze} to provide probe movement instructions for SP acquisition with the help of an external motion tracker. Droste et al.~\cite{droste2020automatic} proposed a behavioral cloning system learned from the action of expert operators to estimate the next probe movement, and it is further extended~\cite{men2022multimodal,men2023gaze} with visual focus that simulates sonographer hand-eye coordination for more accurate guidance. However, these guidance methods are supervised by motion signals and rely on the previous probe motion to infer how to move the probe towards SP. For the detection or guidance methods mentioned, a key overlooked factor is the natural correlation between a freehand scanning 2D frame and its position in the 3D reference frame of the viewed anatomy, which is crucial for navigating within a 3D space during SP searching.

Methods for localizing a 2D US plane in a volumetric data space mainly refer to pose estimation under 3D US~\cite{mohamed2019survey}. With the known 3D spatial information, an accurate 2D to 3D mapping can be achieved for pose estimations~\cite{dou2019agent,namburete2018fully,yang2019fetusmap}. Yeung et al.~\cite{yeung2021learning} attempted to infer the pose of a freehand 2D fetal head image from a 3D US-based network. The estimation is less satisfactory due to the inherent differences between the two imaging techniques. For example, 3D US is normally anatomy-centered, while the structure in a freehand 2D scan can be arbitrary position. To increase cross-modality performance, cycle consistency is later adopted in~\cite{yeung2022adaptive} to link between 3D US slices and freehand images with a minimum manual registration. However, these methods are limited by an effective evaluation metric without the actual 3D location of a 2D US frame, and they are not targeted for SP acquisition guidance.

In this work, we fill the literature gap by combining the two fields: registering the freehand US image to a volumetric atlas and automatically guiding towards SP in 3D anatomy. A uniform model, \textit{Pose-GuideNet}, is proposed to first localize 2D US image without the need for manual registration, and then provide scanning guidance to reach a SP image. The clinical target is head biometry acquisitions, where a sonographer typically searches for the trans-ventricular plane (TVP) and trans-cerebellar plane (TCP) to assess fetal brain development and estimate gestational age. Pose-GuideNet first performs \textit{in-plane alignment} by registering the acquired SP frame in a real-world freehand scan to the biometry view in 3D fetal head atlas. The rest of the scanning frames are then registered through \textit{out-of-plane alignment} with their geometry affinity informed by their anatomical similarity. The method gives a relatively precise measurement by comparing with probe motion data that reflects the true 3D position of the current plane, which also facilitates the new evaluation methods of freehand image localization techniques. Experimental results with both motion-based and image-based evaluations demonstrate that Pose-GuideNet performs accurate pose estimation by preserving the direction and anatomical details of the query 2D image. The estimated transformation towards SP may also drive the real-world deployment of probe displacement in SP acquisition.

\section{Methods}
\begin{figure*}[t]
    \centering
    \includegraphics[width=1.0\textwidth]{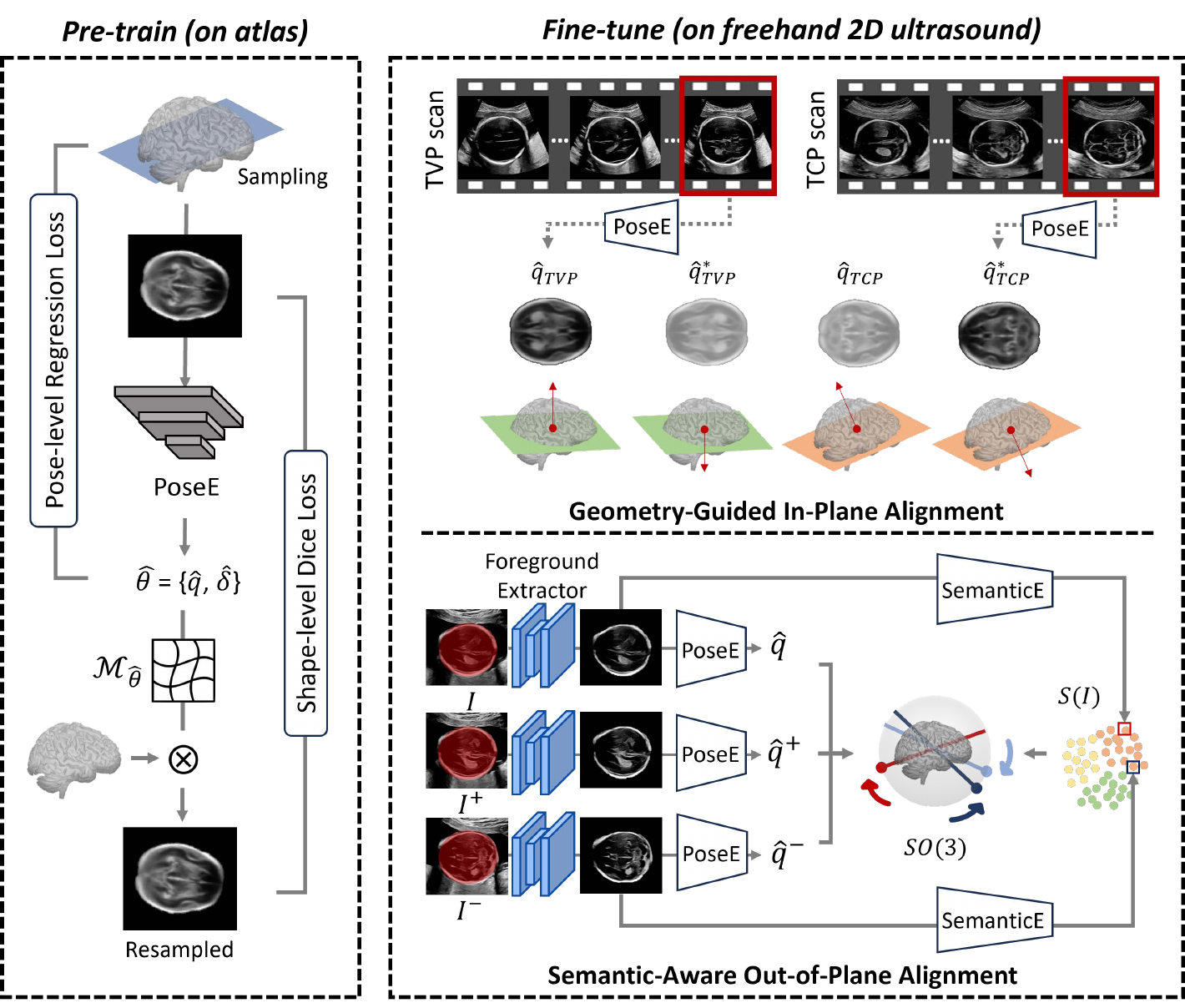}
    \caption{Overview of \textit{Pose-GuideNet} for pose localization of fetal head.}
    \label{fig:framework}
\end{figure*}

Figure~\ref{fig:principle} outlines the process of scanning guidance with \textit{Pose-GuideNet}. The system anticipates the transformation of the current US plane to the pre-defined target viewpoint in 3D anatomy (\textit{i.e.} TVP and TCP in fetal head scan). The transformation can be further converted to probe movement for a human sonographer to manipulate the probe to reach the target plane in freehand 2D scanning. Figure~\ref{fig:framework} depicts the training workflow of the proposed pose localization method. Since the position of 2D US image is unknown, a two-stage learning strategy is considered for Pose-GuideNet: A pose encoder is initially trained on sampled atlas 2D slices to capture the generic mapping between the anatomical feature and its corresponding position in 3D space; It is then fine-tuned on freehand 2D US scans by 1) the \textit{Geometric-guided In-plane Alignment} - supervised alignment of the head standard planes and their directions based on the prior geometric knowledge on the atlas, 2) the \textit{Semantic-aware Out-of-plane Alignment} - unsupervised alignment of the off-plane frames to atlas w.r.t. their anatomical similarity learned by a semantic encoder.

\textbf{Problem Definition.} The principle of the method involves establishing a correspondence between the reference 2D US image and its cross-sectional view in an open-sourced 3D atlas volume~\cite{namburete2023normative}. Given any 2D US image frame $I$ in a freehand fetal head scanning stream, Pose-GuideNet registers the US image by predicting its location $\bm{\theta}=\{\bm{q}, \bm{\delta}\}$ including orientation (in quaternion) $\bm{q}=[w, x, y, z]$ and translation $\bm{\delta} = [\Delta_x, \Delta_y, \Delta_z]$ in 3D brain anatomy. Then the transformation of the US image towards SP is defined as $\bm{q}_I^{sp} = \bm{q}^{*}\bm{q}_{sp}$ (quaternion multiplication) and $\bm{\delta}_I^{sp} = \bm{\delta}_{sp}-\bm{R}(\bm{q}_I^{sp})\cdot\bm{\delta}$, with $\bm{q}_{sp}$ and $\bm{\delta}_{sp}$ representing the orientation and the translation of SP under the same coordinate of the 3D atlas. Here, $\bm{q}^{*}$ denotes the quaternion conjugate and $\bm{R}(\bm{q})$ denotes the $3\times3$ rotation matrix of $\bm{q}$, and $sp$ $\in\{$TVP, TCP$\}$.

\subsection{Pose Localization on 3D Atlas}
Unlike existing methods~\cite{namburete2018fully,yeung2022adaptive} that require the alignment and registration of 3D US volumes to atlas, we directly sample from the atlas volume to learn its positional mapping without introducing the 3D US modality. The backbone of Pose-GuideNet is a pose encoder (PoseE) based on ResNeXt50~\cite{xie2017aggregated}. Different from~\cite{yeung2022adaptive,yeung2021learning} that solely constrain the 3D position, we also incorporate the shape-level similarity loss on the resampled 2D slice generated through a differentiable spatial transformer~\cite{jaderberg2015spatial} to stabilize the training process. This is done by sampling the transformed grid $\mathcal{M}_{\bm{\hat{\theta}}}$ of the estimated parameters $\bm{\hat{\theta}}$ on the atlas volume (denoted as $V$). Since the intensity-based image constraint will easily overwhelm the adaption to US image fine-tuning, dice similarity coefficient (DSC) is used here to preserve the region of interest in the resampled slice. Specifically, the training objective on 3D atlas combines the shape-level dice loss and the pose-level regression loss:
\begin{equation}
    \mathcal{L}_{atlas}(I^{atl}) = 1-\frac{2\sum_k (V \circ\mathcal{M}_{\bm{\hat{\theta}}}(k)) \cdot y(I^{atl}(k))}{\sum_k (V \circ\mathcal{M}_{\bm{\hat{\theta}}}(k))+\sum_iy(I^{atl}(k))} + ||\bm\theta-\bm{\hat{\theta}}||_2,
\end{equation}
where $I^{atl}$ is the input atlas slice at pose $\bm\theta$, $\bm{\hat{\theta}}$ is its predicted pose, and $y(I^{atl}(k))$ is the binary label for $I^{atl}$ at pixel $k$. $\circ$ denotes the grid sampling operation.

\subsection{Cross-Dimension Cross-Modality Alignment on 2D US}
The dimensional gap between 2D US and 3D atlas is one of the main challenges that affects the 2D/3D registration accuracy. To overcome this obstacle, we propose to 1) establish an explicit mapping based on the cross-dimensional shared SP frame, and 2) for the frames with unknown mapping, learn the closeness of pose in 3D from their image-level conformity. The whole process includes the standard plane alignment in a head acquisition scan with its geometric prior and the alignment of its searching planes based on their anatomical features. As a primary concern for head standard plane refinement, we focus on understanding the change in orientation of the head plane without considering its translation. The rotation group defined on the 3D anatomy is isomorphic to \textit{SO(3)}~\cite{huynh2009metrics} on a Riemannian manifold structure, which allows us to establish a geodesic distance representing the most direct path between the orientations of two planes. Here, geodesic distance $d_{G}$ is adopted on the two types of alignments to measure the closeness of 2D US images in the 3D atlas.

Another challenge is the cross-modality difference, and we reduce it by automatic recognition and removal of the unnecessary image background in 2D US through cascading an image localization net~\cite{jaderberg2015spatial} with MedSAM~\cite{ma2024segment}. The extracted foreground of fetal head is further processed by bilinear filtering~\cite{banterle2012low} to produce a smoothed, atlas-like image before pose localization. 

\textbf{Geometry-Guided In-Plane Alignment.} 
To ensure the geometry correctness, the target SP of a scan is aligned to one of the standard viewpoints (TVP or TCP) in atlas, each of which contains two opposite directions with the predefined geometry positions (as shown in the top right of Figure~\ref{fig:framework}). Specifically, given a US SP image $I_{sp}$, the geodesic loss is given by
\begin{equation}
    \mathcal{L}_{us}^{in}(I_{sp})=\alpha d_{G}(\bm{q}_{sp},\bm{\hat{q}}_{sp}) = \arccos(|\!<\bm{q}_{sp},\bm{\hat{q}}_{sp}>\!|)
\end{equation}
that is computed between the SP positional prior $\bm{q}_{sp}$ and its prediction $\bm{\hat{q}}_{sp}$, where $|\!<\cdot,\cdot>\!|$ is the absolute inner product between two normalized quaternions~\cite{mahendran20173d}. Here, $\alpha$ is 0.5 given that the angle on \textit{SO(3)} is half of the angle between their corresponding 3D orientations~\cite{huynh2009metrics}. This in-plane alignment step allows Pose-GuideNet to identify the standard viewpoint, facilitating the geometric error corrections during registration.

\textbf{Semantic-Aware Out-of-Plane Alignment.} 
In a US scan, the searching frames before SP are of diverse angles in the 3D head anatomy, which are essentially spanned across the rotation manifold. Inspired by the unsupervised manifold learning~\cite{iscen2018mining}, we preserve the consistency in US images by contrasting their mappings on the rotation manifold. The manifold nearest neighbors (\textit{i.e.} positive pairs) are selected from consecutive frames in the scan, which are geometrically closest to each other in the scanning sequence. The negative items are randomly selected frames within the same scan. However, given that two timely distinct frames may be visually similar, we also incorporate semantic similarity in negative items to advise their geometric closeness. The semantic similarity is to compare their anatomical characteristics from a semantic encoder (SemanticE) formed by SonoNet~\cite{baumgartner2017sononet}. In particular, the alignment of an out-of-plane frame $I$ can be regularized by a semantic-aware contrastive loss of the quaternion triplet:
\begin{equation}
    \mathcal{L}_{us}^{out}(I) =-\log\frac{\exp(\cos(\alpha d_G(\bm{\hat{q}}, \bm{\hat{q}}^{+}))/\tau)}{\sum_{n=1}^{N}<S(I), S(I_{n}^{-})>\exp(\cos(\alpha d_G(\bm{\hat{q}}, \bm{\hat{q}}_{n}^{-}))/\tau)},\label{eq:cl}
\end{equation}
where $I_{n}^{-}$ is the $n$th negative sample of $I$ with the total number $N$ set to 5. $S(I)$ is the encoded semantic feature, and the temperature $\tau$ is set to 0.8 which controls the uniformity of the embedding distribution~\cite{wang2021understanding}. 

The overall objective for 2D US fine-tuning is to jointly optimize $\mathcal{L}_{us}^{in}$ and $\mathcal{L}_{us}^{out}$ on a batch of US images with in- and out-of-plane collections. 

\section{Experiments}
\subsection{Data and Experimental Settings}
In this study, the 2D US scans were collected as part of the PULSE (Perception Ultrasound by Learning Sonographic Experience)~\cite{drukker2021transforming}. The clinical fetal US scans were conducted on a GE Voluson E8 scanner. The US sequence is selected 10s before the \textit{cine-buffer-corrected} SP and pre-processed to 6 Hz to reduce redundancy. This study received approval from UK Research Ethics Committee. There are 192 scanning sequences for the 2\textsuperscript{nd} trimester fetal head biometry acquisition, resulting in 10810 valid image planes. The training and test split is 7428/3382. Among all test planes, 2081 are for TVP scans and 1301 for TCP. At each plane, the synchronized probe orientation $\bm{\tilde{q}}$ is recorded by an IMU motion sensor, which is used for evaluation purpose. The 3D fetal head atlas used is open-sourced~\cite{namburete2023normative}. 1059 US brain volumes collected from 899 fetuses were used to generate the fetal head atlases for different gestational ages. Here, we use the atlas at 20 week pregnancy as the median description of the fetal head in the 2\textsuperscript{nd} trimester. To better align with the atlas slice, the US video frame was cropped to 224×288 and resized to 160×160 while keeping the aspect ratio. The experiments were run with PyTorch 1.10.1 on a 32GB NVIDIA Tesla V100 GPU. Pose-GuideNet is pre-trained and fine-tuned for 800 and 100 epochs respectively, using Adam optimizer with a batch size of 8 and a learning rate of 1e-4. To prevent overfitting, the last linear layer in PoseE is frozen during fine-tuning.

\begin{table}[]
\centering
\caption{Quantitative results of the estimated plane in terms of motion transformation towards SP and image-level similarity.}
\resizebox{\columnwidth}{!}{
\begin{tabular}{c|l|cccc}
\hline
\multirow{3}{*}{}             & \multirow{3}{*}{Architecture} & \multicolumn{4}{c}{TVP}                                                                                  \\ \cline{3-6} 
                              &                               & \multicolumn{1}{c|}{Motion}         & \multicolumn{3}{c}{Image}                                          \\ \cline{3-6} 
                              &                               & \multicolumn{1}{c|}{KL Divergence $\downarrow$}  & Dice(\%) $\uparrow$          & NCC $\uparrow$                & MS-SSIM $\downarrow$ \\ \hline
 \multirow{5}{*}{Sensorless}              & Random plane                  & \multicolumn{1}{c|}{19.89}          & 43.52          & 0.169$\pm$0.03          & 0.614$\pm$0.07                  \\ \cline{2-6}
             & Yeung et al.                  & \multicolumn{1}{c|}{18.89}          & 55.03          & 0.271$\pm$0.07          & 0.556$\pm$0.08                  \\
                              & Ours (\textit{pre-train})               & \multicolumn{1}{c|}{18.17}          & 65.47          & 0.277$\pm$0.09          & 0.537$\pm$0.10                  \\
                              & Ours (\textit{ft. Geo.})            & \multicolumn{1}{c|}{18.37}          & 65.59          & 0.306$\pm$0.08          & 0.530$\pm$0.07                  \\
                              & Ours (\textit{ft. Geo. Sem.})       & \multicolumn{1}{c|}{\textbf{17.33}} & \textbf{66.91} & \textbf{0.315$\pm$0.09} & \textbf{0.529$\pm$0.09}         \\ \hline
\multirow{2}{*}{Sensored}   & US-GuideNet                   & \multicolumn{1}{c|}{15.01}          & -              & -                   & -                           \\
                              & Multimodal-GuideNet           & \multicolumn{1}{c|}{\textbf{14.79}} & -              & -                   & -                           \\ \hline \hline
\multirow{3}{*}{}             & \multirow{3}{*}{Architecture} & \multicolumn{4}{c}{TCP}                                                                                  \\ \cline{3-6} 
                              &                               & \multicolumn{1}{c|}{Motion}         & \multicolumn{3}{c}{Image}                                          \\ \cline{3-6} 
                              &                               & \multicolumn{1}{c|}{KL Divergence $\downarrow$}  & Dice(\%) $\uparrow$         & NCC $\uparrow$                & MS-SSIM $\downarrow$\\ \hline
       \multirow{5}{*}{Sensorless}                       & Random plane                  & \multicolumn{1}{c|}{20.32}          & 44.89          & 0.183$\pm$0.03          & 0.619$\pm$0.06                  \\ \cline{2-6}
 & Yeung et al.                  & \multicolumn{1}{c|}{20.03}          & 57.11          & 0.308$\pm$0.10          & 0.536$\pm$0.08                  \\
                              & Ours (\textit{pre-train})               & \multicolumn{1}{c|}{19.13}          & 69.11          & 0.312$\pm$0.05          & 0.532$\pm$0.05                  \\
                              & Ours (\textit{ft. Geo.})            & \multicolumn{1}{c|}{16.42}          & 67.19          & 0.341$\pm$0.08          & 0.521$\pm$0.07                  \\
                              & Ours (\textit{ft. Geo. Sem.})       & \multicolumn{1}{c|}{\textbf{15.47}}          & \textbf{70.12} & \textbf{0.353$\pm$0.08} & \textbf{0.511$\pm$0.07}         \\ \hline
\multirow{2}{*}{Sensored}   & US-GuideNet                   & \multicolumn{1}{c|}{15.27} & -              & -                   & -                           \\
                              & Multimodal-GuideNet           & \multicolumn{1}{c|}{\textbf{14.62}}          & -              & -                   & -                           \\ \hline
\end{tabular}\label{tab:result}}
\end{table}

\subsection{Evaluations with Motion-level Correspondence}
A clinical goal for 2D US image localization is to find the biometry plane for fetal measurements. Because of the inherent correspondence between the probe manipulation and the field of view change in the scanned anatomy~\cite{ihnatsenka2010ultrasound}, we utilize the probe rotation towards its reached SP, \textit{i.e.} $\bm{\tilde{q}}_{I}^{sp}=\bm{\tilde{q}}^*\bm{\tilde{q}}_{sp}$, for evaluation of the predicted transformation between the current plane and its expected standard plane, \textit{i.e.} $\bm{\hat{q}}_{I}^{sp}=\bm{\hat{q}}^*\bm{q}_{sp}$. Since the probe and the 3D fetal head positioning are operated under different coordinate systems, their angular or positional distances are not directly comparable. We thus evaluate on the distributions of these two rotation dynamics in a scan using the statistical metric of KL Divergence. Other than the variations of our method, we compared with random plane assignment and registration-based approach~\cite{yeung2021learning} with atlas pre-training, and US guidance methods \textit{US-GuideNet}~\cite{droste2020automatic} and \textit{Multimodal-GuideNet}~\cite{men2023gaze} based on probe motion prediction towards SP. These probe motion-based methods are considered \textit{sensored} as they predict under a behavioral cloning strategy given the previous step from a motion sensor. 

The results of different architectures are given in the \textit{Motion} column of Table~\ref{tab:result}. On these two types of head biometry scans, Pose-GuideNet consistently improves under the \textit{sensorless} method by gradually incorporating in-plane geometric (\textit{ft. Geo.}) and out-of-plane semantic (\textit{ft. Geo. Sem.}) alignments, which reduce the motion error close to the baselines using extra sensor(s). Practically, it is more challenging to locate TCP than TVP with the anatomical structure of cerebellum much smaller in the brain. TCP is thus less likely to be sampled during pre-training with a larger localization error than TVP (20.03$>$18.89 and 19.13$>$18.17). However, by fine-tuning with geometric-guided alignment, Pose-GuideNet can greatly improve cross-dimensional registration by reducing the motion difference from 19.13 to 16.42. This is because the key plane serves as a common baseline that establishes a spatial correspondence between the 2D US and 3D anatomy. This correspondence is necessary for accurate mapping especially for few-shot planes such as TCP. 

\begin{figure*}[t]
    \centering
    \includegraphics[width=1.0\textwidth]{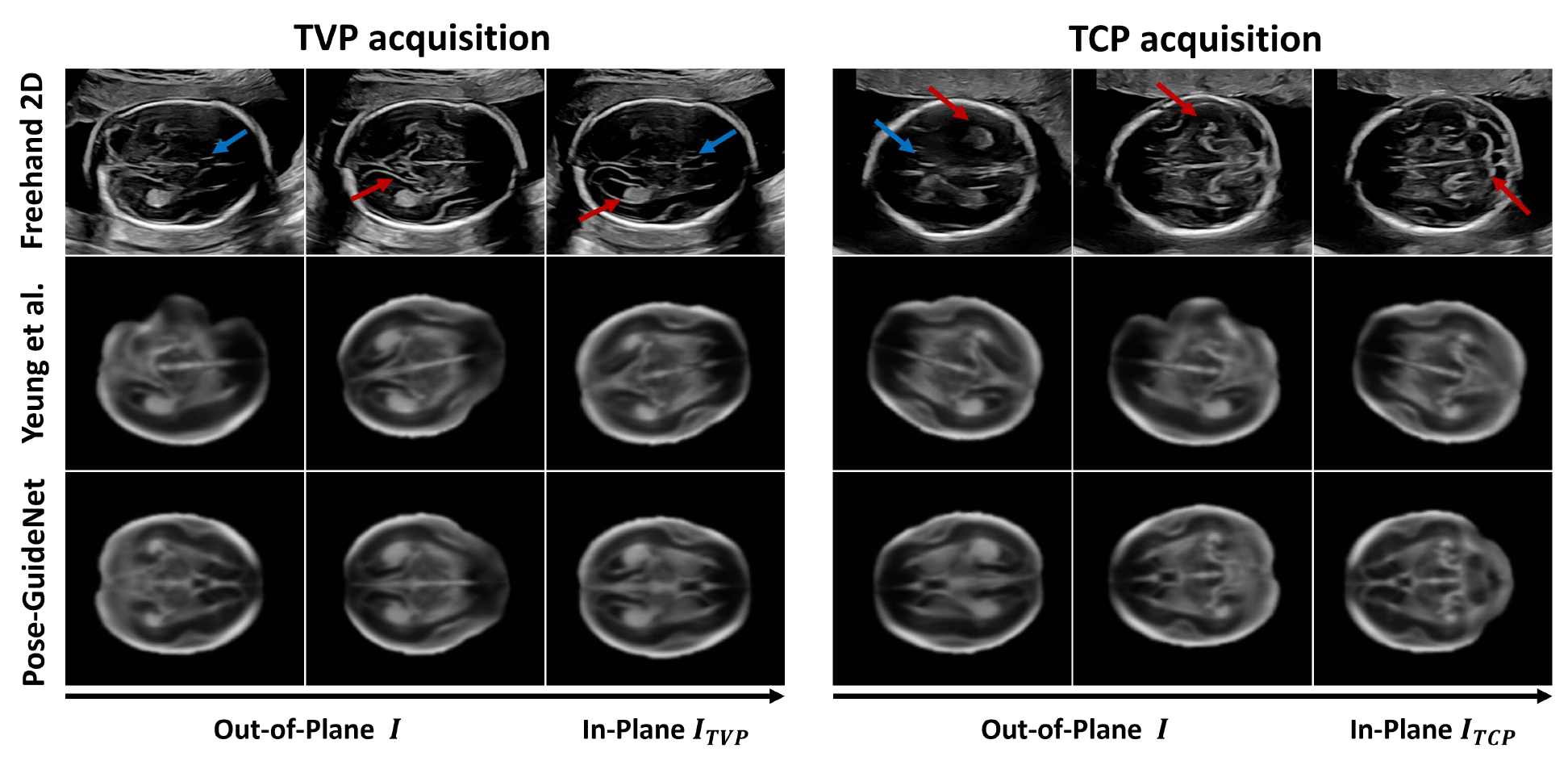}
    \caption{The retrieved atlas planes of two example 2D head biometry acquisitions.}
    \label{fig:visual}
\end{figure*}

\subsection{Evaluations with Image-level Correspondence}
The visual similarity between the queried 2D US image and the retrieved plane from 3D anatomy is also compared in Table~\ref{tab:result}. Dice score, Normalized Correlation Coefficient (NCC), and Multi-Scale Structural Similarity Index Measure (MS-SSIM) are used as image similarity metrics, where MS-SSIM is more comprehensive than SSIM to tolerate scale variations of US images. From the results, Pose-GuideNet generally shows consistent performance across different metrics for both planes. A drop of dice score for TCP is observed when including the geometric constraint (69.11\% vs. 67.19\%). This may be due to that in atlas, the occipital bone ending around the cerebellum can be greatly deformed within a small angle change~\cite{namburete2023normative} (as in supplementary material), which makes it sensitive to the segmentation-based measurement. Figure~\ref{fig:visual} shows the result of retrieved atlas planes from in- and out-of-plane frames in two head scanning sequences. The anatomical structures (as highlighted by colored arrows) and head directions are both aligned well with the input 2D image under Pose-GuideNet. Some key structures, such as cavum septum pellucidum (CSP) highlighted in blue arrows, are challenging to observe due to the acoustic shadow caused by ultrasound artifacts. However, Pose-GuideNet can better estimate them by learning from the appearance of a neighbor frame with a different luminance condition.

\section{Conclusion}
We propose Pose-GuideNet which estimates the 3D pose from freehand 2D US image and its transformation towards SP in 3D anatomy, for the guidance of fetal head biometry acquisition. The approach registers US image to a 3D atlas volume using its prior knowledge of SP geometry as a baseline mapping, and using the anatomical similarity to consistently align the rest of off-plane frames in the scan. Extensive experiments show that the predicted plane is highly aligned with the input US image in terms of both anatomical structures and direction. Note that the Pose-GuideNet is a sensor-free method that can stand alone for fetal head pose estimation. As a future step, geometric correspondence can be incorporated that converts the transformation in 3D anatomy to probe motion in real-world deployment. This will enable a multi-functional guidance system for trainee sonographers, by providing not only a reference plane in 3D but also the decisive motion in hand towards the target plane.

\begin{credits}
\subsubsection{\ackname} We acknowledge UKRI grant reference (EP/X040186/1), EPSRC grant (EP/T028572/1), and ERC grant (ERC-ADG-2015 694581, project PULSE). 

\subsubsection{\discintname}
The authors have no competing interests in the paper as required by the publisher.
\end{credits}

%
%
%
\bibliographystyle{splncs04}
\bibliography{Paper}

\end{document}